\documentclass[runningheads]{llncs}
\usepackage[T1]{fontenc}
\usepackage{graphicx}
\usepackage{amsmath}
\usepackage{multirow}
\usepackage{amssymb}
\usepackage[colorlinks=true, linkcolor=black, urlcolor=blue]{hyperref}
\usepackage{orcidlink}

\begin{document}
\title{Adaptive k-space Radial Sampling for Cardiac MRI with Reinforcement Learning}

\author{Ruru Xu, Ilkay Oksuz\orcidlink{0000-0001-6478-0534}}
\authorrunning{Ruru Xu et al.}
\institute{Computer Engineering Department, Istanbul Technical University, Istanbul, Turkey
\email{xu21@itu.edu.tr}\\}

\maketitle              
\begin{abstract}
Accelerated Magnetic Resonance Imaging (MRI) requires careful optimization of k-space sampling patterns to balance acquisition speed and image quality. While recent advances in deep learning have shown promise in optimizing Cartesian sampling, the potential of reinforcement learning (RL) for non-Cartesian trajectory optimization remains largely unexplored. In this work, we present a novel RL framework for optimizing radial sampling trajectories in cardiac MRI. Our approach features a dual-branch architecture that jointly processes k-space and image-domain information, incorporating a cross-attention fusion mechanism to facilitate effective information exchange between domains. The framework employs an anatomically-aware reward design and a golden-ratio sampling strategy to ensure uniform k-space coverage while preserving cardiac structural details. Experimental results demonstrate that our method effectively learns optimal radial sampling strategies across multiple acceleration factors, achieving improved reconstruction quality compared to conventional approaches.  Code available: \url{https://github.com/Ruru-Xu/RL-kspace-Radial-Sampling} 

\keywords{Cardiac MRI  \and Radial Sampling \and MRI Reconstruction \and Reinforcement Learning.}
\end{abstract}

\section{Introduction}
Magnetic Resonance Imaging (MRI) has revolutionized medical diagnosis through its ability to provide high-quality soft tissue visualization without ionizing radiation. The acquisition process in MRI involves sampling the k-space (frequency domain) data, where the sampling strategy directly impacts both image quality and scan time. While various acceleration techniques like parallel imaging \cite{griswold2002generalized}, compressed sensing \cite{lustig2007sparse}, and more recent deep learning approaches \cite{jin2019self} have been developed, determining optimal k-space sampling patterns remains a fundamental challenge in MRI acceleration.

Reinforcement learning (RL) has recently emerged as a promising approach for optimizing k-space sampling strategies, offering data-driven solutions that can adapt to different imaging scenarios. While existing RL-based methods have shown success in optimizing Cartesian sampling trajectories \cite{pineda2020active}, the potential of RL in non-Cartesian sampling remains largely unexplored. Radial sampling \cite{winkelmann2006optimal}, in particular, offers inherent advantages such as motion robustness and efficient k-space center coverage, making it especially suitable for cardiac imaging applications. However, designing optimal radial sampling patterns presents unique challenges due to its non-uniform k-space coverage and complex trajectory optimization.

In this work, we explore the application of RL to radial sampling optimization, introducing a novel framework that jointly considers both k-space and image-domain information. Our key contributions include:
\begin{itemize}
    \item A dual-branch architecture that jointly processes k-space and image-domain features, enabling comprehensive information utilization from both domains
    \item A cross-attention fusion mechanism that facilitates effective information exchange between k-space and image representations
    \item A mathematically principled golden-ratio sampling strategy that ensures uniform k-space coverage
    \item An anatomically-aware reward design that balances global image quality and cardiac region fidelity.
\end{itemize}

The proposed method represents one of the first attempts to optimize radial sampling trajectories through reinforcement learning in cardiac MRI. By combining dual-domain feature processing with golden-ratio modulated sampling, our framework provides a principled approach to balance k-space coverage and reconstruction quality, providing insights for future development of learning-based non-Cartesian sampling strategies.

\section{Related Work}

Deep learning has revolutionized MRI acceleration through learned k-space sampling optimization. Recent work by \cite{alkan2024autosamp} introduced AutoSamp, employing variational information maximization for optimizing 3D MRI sampling trajectories. \cite{yen2024adaptive} developed an adaptive sampling strategy for rapid pathology prediction, demonstrating superior performance in various clinical applications. The fundamental work by \cite{bahadir2019learning}\cite{wang2025towards}\cite{lyu2025state} established a learning-based framework for optimizing k-space sampling patterns.
The emergence of reinforcement learning (RL) has opened new avenues for dynamic k-space sampling optimization. \cite{xu2024segmentation} demonstrated the effectiveness of RL in cardiac MRI through a segmentation-aware approach that maintains diagnostic accuracy while improving reconstruction efficiency. Meanwhile, \cite{yang2024attention} showed how attention mechanisms can enhance reconstruction quality in specialized applications like CEST MRI.

Recent advances in non-Cartesian sampling have shown particular promise in motion-sensitive applications. \cite{radhakrishna2023jointly} proposed a novel approach for jointly learning non-Cartesian k-space trajectories and reconstruction networks for both 2D and 3D MR imaging. This builds upon earlier work by \cite{weiss2020joint}, who demonstrated the advantages of joint learning for Cartesian undersampling and reconstruction in accelerated MRI.
We introduce a learning-based framework that uniquely combines the advantages of reinforcement learning and radial sampling for robust MRI acquisition, addressing the limitations of existing approaches while maintaining high reconstruction quality.

\section{Method}

\subsection{Network Architecture}
\begin{figure}[tbp]
    \centering
    \includegraphics[width=1\linewidth]{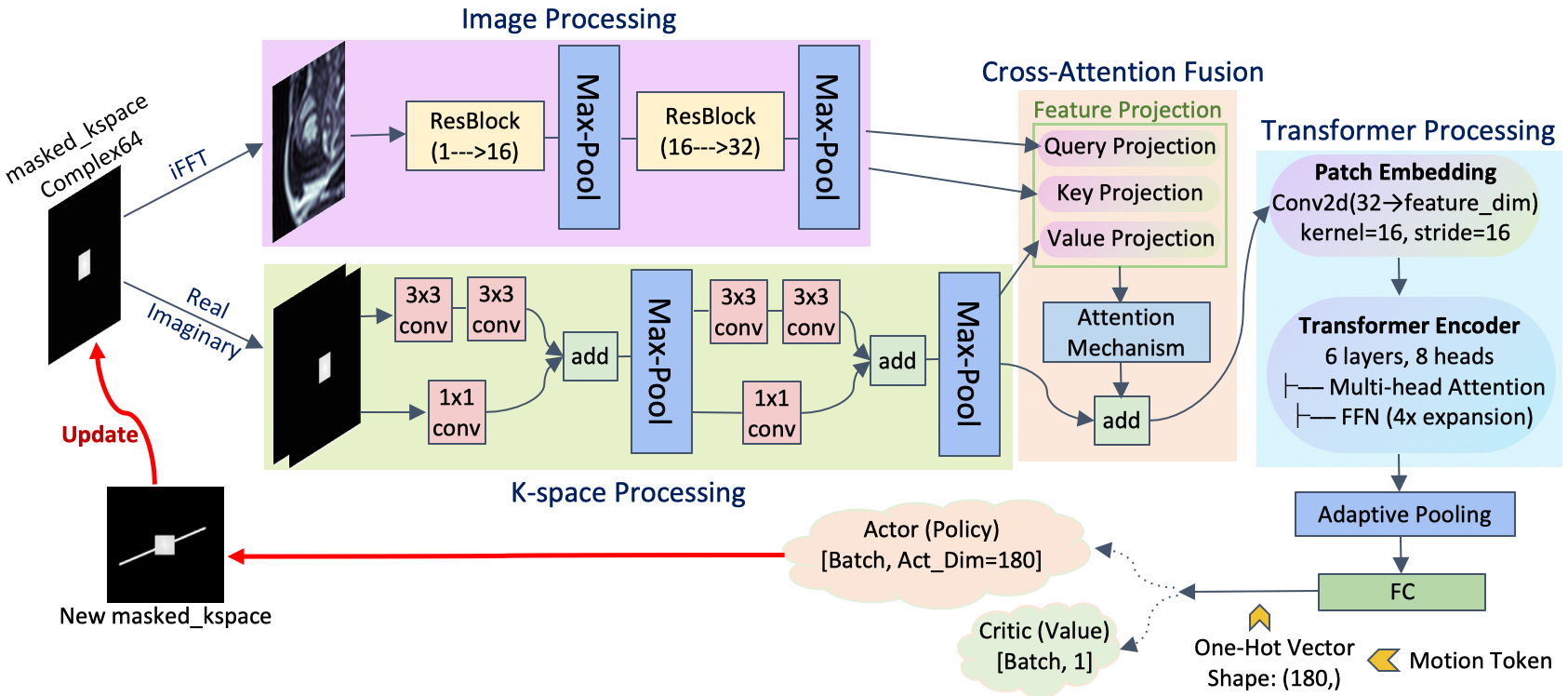}
    \caption{The proposed dual-branch architecture with cross-attention fusion for k-space sampling optimization. The network processes complex-valued k-space data and image-domain features in parallel, integrating them through a transformer-based framework for policy learning.}
    \label{fig:network}
\end{figure}

We propose a dual-branch actor-critic architecture for optimizing k-space sampling trajectories in MRI reconstruction. As illustrated in Figure \ref{fig:network}, our framework processes partially sampled k-space data through parallel branches that operate in both k-space and image domains, initialized with a 32×32 central calibration region. The k-space branch processes complex-valued data through cascaded residual blocks with progressive channel expansion, while the image branch simultaneously analyzes the inverse Fourier transformed data to leverage spatial coherence. 

To effectively integrate information across domains, we introduce a cross-attention fusion module that dynamically weights features from both branches through scaled dot-product attention. The fused features are processed by a transformer encoder with multi-head attention and patch embedding, employing pre-norm configuration and dropout regularization to ensure stable training. The network outputs both policy and value estimates through separate heads: the policy head predicts sampling angles through fully-connected layers with progressive dropout, while the value head guides policy optimization through critic learning. 

\subsection{Radial Sampling Mask Generation}
\begin{figure}[h]
    \centering
    \includegraphics[width=1\linewidth]{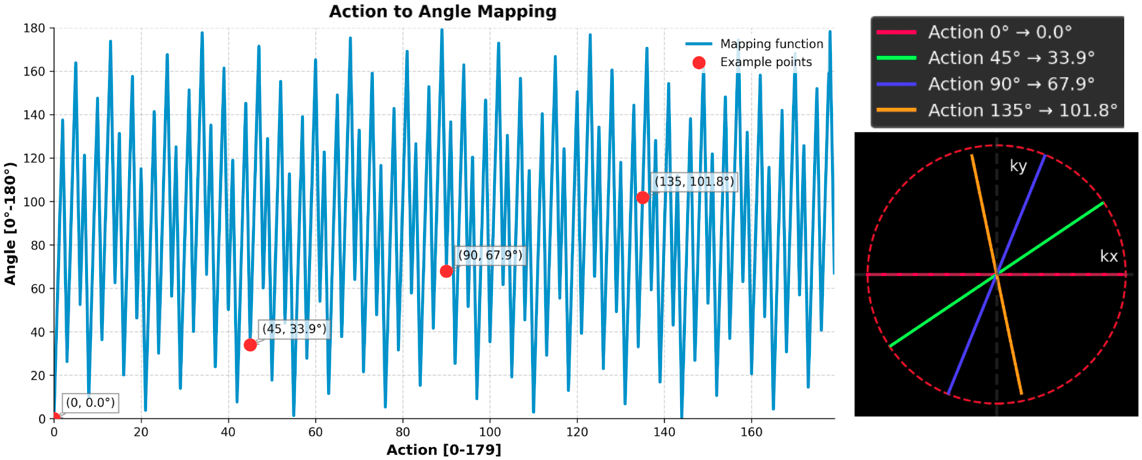}
    \caption{Visualization of the action-to-angle mapping mechanism in k-space. The continuous mapping function $\mathcal{T}: \mathbb{A} \rightarrow \Theta$ transforms discrete actions to angular space via golden ratio modulation. Right panel demonstrates the resulting k-space coverage patterns, illustrating the uniform angular distribution property.}
    \label{fig:sampling}
\end{figure}

We propose an optimal k-space sampling strategy based on golden-ratio modulated angular distribution. Given action space $\mathbb{A} = \{0,\ldots,179\}$, the angular mapping function $\mathcal{T}: \mathbb{A} \rightarrow \Theta$ is defined as:

\begin{equation}
    \mathcal{T}(a) = (a\varphi_g) \bmod \pi, \quad a \in \mathbb{A}
    \label{eq:angle}
\end{equation}

where $\varphi_g = 137.508 / 2 = 68.754°$ represents the optimal golden angle \cite{feng2022golden} increment for uniform k-space coverage. The binary sampling mask $\mathcal{M}: \Theta \rightarrow \{0,1\}^{N \times N}$ is formulated through composition of operators:

\begin{equation}
    \mathcal{M}(\theta) = \mathcal{B}(\mathcal{R}_{\theta}) \circ \mathcal{C}_r
    \label{eq:mask}
\end{equation}

Here, $\mathcal{R}_{\theta}$ denotes the radial trajectory operator, $\mathcal{B}$ represents binary discretization, and $\mathcal{C}_r$ defines the circular support constraint. For multi-line acquisition with stride $n$, the composite mask $\mathcal{M}_n$ is constructed as:

\begin{equation}
    \mathcal{M}_n(\theta) = \bigvee_{i=0}^{n-1} \mathcal{M}(\theta + i\delta\theta), \quad \delta\theta = \frac{\pi}{4n}
    \label{eq:multiline}
\end{equation}

The temporal evolution of the sampling pattern $\mathcal{A}_t$ follows:

\begin{equation}
    \mathcal{A}_t = \sup(\mathcal{A}_{t-1}, \mathcal{M}_n(\mathcal{T}(a_t)))
    \label{eq:accumulation}
\end{equation}

This formulation guarantees optimal k-space coverage through golden angle modulation while maintaining sampling efficiency and uniform spatial frequency distribution.

\subsection{Reward Function}
We propose an anatomically-aware reward function that optimizes both global reconstruction fidelity and region-specific cardiac details through a dynamic weighting mechanism. The temporal reward $R_t$ is formulated as:

\begin{equation}
    R_t = \lambda_g(\alpha_t \nabla\mathcal{S}_g + \beta_t \nabla\mathcal{S}_h) - \lambda_h\nabla\mathcal{L}_h
    \label{eq:reward}
\end{equation}

where $\nabla\mathcal{S}_g$, $\nabla\mathcal{S}_h$ denote the temporal gradients of global and heart-specific structural similarity indices respectively, and $\nabla\mathcal{L}_h$ represents the cardiac region reconstruction loss gradient. The adaptive weighting coefficients $(\alpha_t, \beta_t)$ are determined by:

\begin{equation}
    (\alpha_t, \beta_t) = \begin{cases}
        (\gamma_b, 1-\gamma_b) & \text{if } \|\mathcal{S}_g - \mathcal{S}_h\|_1 < \epsilon \\
        (\max(\gamma_{min}, \gamma_b e^{-\eta t/T}), 1-\alpha_t) & \text{otherwise}
    \end{cases}
    \label{eq:weight}
\end{equation}

The cardiac-specific metrics are computed through masked evaluation:

\begin{equation}
    \mathcal{S}_h = \mathcal{F}_{ssim}(\mathbf{x}_{\Omega}, \hat{\mathbf{x}}_{\Omega}), \quad 
    \mathcal{L}_h = \frac{\|\mathbf{x}_{\Omega} - \hat{\mathbf{x}}_{\Omega}\|_2^2}{\|\mathbf{x}_{\Omega}\|_{\infty}^2}
    \label{eq:metrics}
\end{equation}

where $\Omega$ represents the cardiac region of interest. This formulation ensures progressive emphasis on cardiac structure preservation while maintaining global reconstruction quality through exponentially decaying global attention ($\gamma_b=0.4$, $\gamma_{min}=0.1$, $\eta=0.3$, $\lambda_g=0.6$, $\lambda_h=0.4$).

\section{Experiments and Results Analysis}

\noindent\textbf{Datasets:} We utilize the Automated Cardiac Diagnosis Challenge (ACDC) dataset \cite{bernard2018deep}, which contains cardiac MRI sequences from 150 patients with various pathologies. We divide it into training (94 patients, 1,783 slices), validation (20 patients, 367 slices), and test (36 patients, 692 slices) sets. Our preprocessing pipeline includes spatial standardization, slice-wise intensity normalization, and k-space transformation.

\noindent\textbf{Implementation Details:} Our experiments were conducted on NVIDIA GV100 (training batch size: 30, validation batch size: 160) and RTX 3090 (training batch size: 10, validation batch size: 120) GPUs, evaluating acceleration factors of $4\times$, $8\times$, and $12\times$. The network was trained using PPO algorithm with the following configurations: (1) Optimization: AdamW optimizer (lr=$4.5 \times 10^{-4}$, weight decay=$10^{-4}$) with step-wise learning rate scheduling (step size=1000, decay factor=0.98); (2) PPO parameters: discount factor $\gamma=0.99$, GAE-$\lambda=0.9$, entropy coefficient=0.08, value function coefficient=0.3, and gradient clipping threshold=0.5; (3) Training protocol: 6 update epochs per iteration with evaluation every 20 intervals. Considering the instability at the beginning of training, the reward between the two steps is not obvious. We first used multi-line acquisition; that is, one action gets two sampling lines and then we switched to a single-line acquisition.

\begin{table}[htbp]
\centering
\caption{Performance comparison of different sampling strategies for accelerated cardiac MRI reconstruction. The initial sampling uses a fixed 32×32 center radial pattern, while subsequent methods employ different reward functions at various acceleration factors. The Random baseline generates radial sampling lines through random angle selection, $\Delta$SSIM uses temporal SSIM difference as reward, and Region-based method employs our proposed anatomically-aware reward that balances global and cardiac-specific reconstruction quality ($0.6*(\alpha*\Delta SSIM\_global + \beta*\Delta SSIM\_heart) - 0.4*\Delta MSE\_heart$).}
\renewcommand{\arraystretch}{1.0}  
\setlength{\tabcolsep}{5.8pt}  
\begin{tabular}{|c|l|cc|cc|c|}
\hline
\multirow{2}{*}{\begin{tabular}[c]{@{}c@{}}Acceleration\\ Factor\end{tabular}} 
& \multirow{2}{*}{Method (Reward)} 
& \multicolumn{2}{c|}{Global} 
& \multicolumn{2}{c|}{Heart Region} 
& \multirow{2}{*}{\begin{tabular}[c]{@{}c@{}}Dice\\ Score\end{tabular}} \\ \cline{3-6}
&  & SSIM & PSNR & SSIM & PSNR &  \\ \hline
64× & Initial (32×32) & 0.6207 & 21.927 & 0.6565 & 22.071 & 0.6548 \\ \hline
\multirow{3}{*}{12×} 
& Random                  & 0.6738 & 23.599 & 0.7335 & 23.949 & 0.7884 \\
& $\Delta$ SSIM            & \textbf{0.7214} & 24.293 & 0.7311 & 24.325 & 0.7667 \\
& Region-based    & 0.7130 & 24.426 & \textbf{0.7470} & 24.631 & \textbf{0.8020} \\ \hline
\multirow{3}{*}{8×} 
& Random                  & 0.7136 & 24.719 & 0.7764 & 25.290 & 0.8211 \\
& $\Delta$ SSIM            & \textbf{0.7634} & 25.443 & 0.7716 & 25.538 & 0.8031 \\
& Region-based    & 0.7524 & 25.632 & \textbf{0.7908} & 26.103 & \textbf{0.8302} \\ \hline
\multirow{3}{*}{4×} 
& Random                  & 0.8226 & 28.664 & 0.8702 & 29.626 & 0.8366 \\
& $\Delta$ SSIM            & \textbf{0.8619} & 30.003 & 0.8903 & 30.857 & 0.8702 \\
& Region-based    & 0.8601 & 29.991 & \textbf{0.8915} & 30.960 & \textbf{0.8718} \\ \hline
\end{tabular}
\label{tab:result}
\end{table}

\subsection{Quantitative Results}
We conduct comprehensive evaluations on the ACDC dataset to validate our proposed reward functions against random radial sampling baseline. Starting from an initial 32×32 center-focused sampling pattern (64× acceleration), we explore different reward strategies across various acceleration factors (4×, 8×, 12×). We test the dice score with nnU-Net \cite{isensee2021nnu}. The results demonstrate three key advantages:

The quantitative results in Table \ref{tab:result} show that both our proposed rewards outperform the random radial sampling baseline. At high acceleration (12×), while the $\Delta$SSIM approach achieves better global reconstruction, our Region-based method demonstrates superior heart region structure preservation. This balanced performance is particularly significant for radial sampling's inherent central-dense nature.

Both reward functions maintain consistent effectiveness across acceleration factors. At moderate acceleration (8×), the Region-based method achieves optimal heart-region metrics while $\Delta$SSIM maintains better global quality. This complementary performance demonstrates the effective balance between global and local reconstruction quality.

Most notably, our Region-based reward demonstrates superior diagnostic accuracy through consistently higher Dice scores (0.8020, 0.8302, and 0.8718 for 12×, 8×, and 4× respectively) compared to both baselines. This improvement in structural preservation is crucial for clinical applications where diagnostic reliability is paramount.

\begin{figure}[t]
    \centering
    \includegraphics[width=1\linewidth]{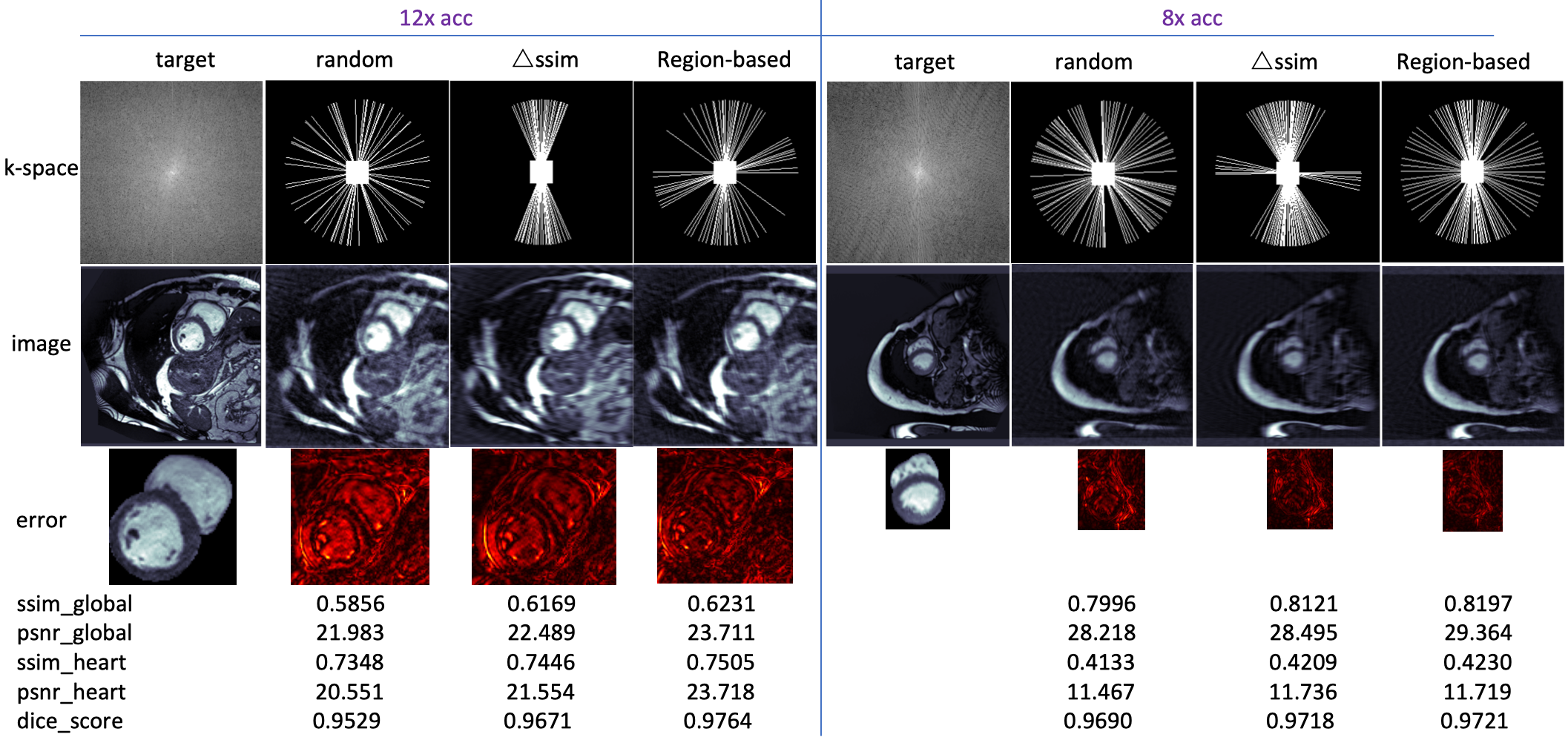}
    \caption{Visual comparison of different sampling strategies at 12× and 8× acceleration factors. Top row: k-space sampling patterns showing the distribution of acquired measurements. Middle row: reconstructed cardiac MR images demonstrating the preservation of anatomical structures. Bottom row: error maps highlighting the spatial distribution of reconstruction errors, with our Region-based method showing reduced errors in cardiac regions. Quantitative metrics below each column indicate superior performance of the proposed approach in both global and cardiac-specific measures.}
    \label{fig:visulize}
\end{figure}

\subsection{Qualitative Results}
The visual comparisons in Fig. \ref{fig:visulize} provide deeper insights into our method's advantages through three key aspects: 
Our region-based approach generates anatomically informed sampling patterns that effectively balance between central and peripheral k-space regions. The sampling patterns demonstrate improved structural organization compared to random and $\Delta$SSIM methods, particularly evident at both 12× and 8× acceleration factors. This sophisticated sampling strategy enables better preservation of both low and high-frequency components essential for diagnostic quality.

The improved sampling pattern translates directly to superior image quality, as evidenced by the reconstructed images. At 12× acceleration, our Region-based method achieves better preservation of heart region structures compared to random sampling. Similar improvements are observed at 8× acceleration, where our method maintains higher fidelity in heart region details.

Analysis of the error maps reveals a significant reduction in reconstruction errors within cardiac regions. At 12× acceleration, our method achieves a dice score of 0.9764 compared to 0.9529 for random sampling, indicating better preservation of cardiac structures. The error distribution pattern demonstrates improved uniformity, particularly in diagnostically critical regions, with global SSIM values of 0.6231 at 12× and 0.8197 at 8× acceleration.

\subsection{Ablation Study}

\begin{table}[tbp]
\centering
\caption{Ablation study comparing different architectural variants of our model at 12× acceleration. V1: k-space features only; V2: image domain features only; V3: simple feature concatenation replacing cross-attention fusion; V4: CNN replacing transformer encoder. Performance is evaluated using SSIM and PSNR (both global and heart-specific) along with segmentation dice score.}
\renewcommand{\arraystretch}{1.0}
\setlength{\tabcolsep}{12pt}
\begin{tabular}{|c|cc|cc|c|}
\hline
Method & \multicolumn{2}{c|}{Global} & \multicolumn{2}{c|}{Heart Region} & dice \\
\cline{2-5}
 & SSIM & PSNR & SSIM & PSNR & score \\
\hline
V1 & 0.7145 & 24.051 & 0.7228 & 24.058 & 0.7447 \\

V2 & 0.7108 & 24.010 & 0.7235 & 24.031 & 0.7549 \\

V3 & 0.7090 & 24.057 & 0.7259 & 24.133 & 0.7564 \\

V4 & 0.7169 & 24.095 & 0.7242 & 24.092 & 0.7445 \\

Ours & \textbf{0.7214} & \textbf{24.293} & \textbf{0.7311} & \textbf{24.325} & \textbf{0.7667} \\
\hline
\end{tabular}
\label{tab:ablation}
\end{table}

To validate our architectural design choices, we conduct ablation studies at 12× acceleration through four variants: V1 (k-space features only), V2 (image domain features only), V3 (simple concatenation instead of cross-attention fusion), and V4 (CNN replacing transformer encoder). As shown in Table \ref{tab:ablation}, our complete model consistently outperforms all variants across metrics. The dual-domain processing proves essential, with single-domain variants (V1, V2) showing inferior performance in both global and heart-region metrics. The effectiveness of cross-attention fusion is demonstrated by V3's reduced global SSIM, while the transformer's importance is validated by V4's lower dice score. These results confirm that each component - dual-domain processing, cross-attention fusion, and transformer-based modeling - contributes significantly to the model's optimal performance in cardiac MRI reconstruction.

\section{Conclusion}
In this work, we presented a novel reinforcement learning framework for optimizing radial sampling trajectories in accelerated cardiac MRI. Our method demonstrates that intelligent sampling pattern selection, guided by anatomically-aware rewards, can significantly improve reconstruction quality while maintaining high acceleration factors. The experimental results validate our approach's effectiveness in learning optimal radial sampling strategies that balance global k-space coverage with cardiac structure preservation.

The success of this approach establishes data-driven sampling optimization as a promising direction for accelerated MRI acquisition. By incorporating anatomical knowledge into the sampling strategy, our method achieves both efficient k-space coverage and reliable cardiac detail preservation. This work opens new possibilities for developing non-cartesian sampling strategies with Reinforcement Learning, potentially benefiting clinical workflows where both speed and diagnostic accuracy are crucial.


\begin{credits}
\subsubsection{\ackname} 
This paper has been benefitted from the 2232 International Fellowship for Outstanding Researchers Program of TUBITAK (Project No: 118C353). The paper also benefited from the TUBITAK bilateral research grant (Project No: 124N419) and the ITU BAP research funds (Project ID: 47296). However, the entire responsibility of the thesis belongs to the owner. The financial support received from TUBITAK does not mean that the content of the thesis is approved in a scientific sense by TUBITAK.
\end{credits}
%
%
%
%

\newpage

\section*{Supplementary Material}
\appendix 
\renewcommand{\thesection}{\Alph{section}} 

\section{Golden-Ratio Modulated Sampling Strategy}

\subsubsection{Mathematical Foundation and Theoretical Framework}

The classical golden angle sampling paradigm employs $\theta_{\text{standard}} = 137.508° = 2\pi(2-\phi)$ where $\phi = \frac{1+\sqrt{5}}{2}$ denotes the golden ratio \cite{winkelmann2006optimal,chan2009temporal}. This angular increment minimizes the Weyl criterion for uniform distribution on the unit circle, achieving optimal discrepancy bounds $D_N = O(\log N/N)$ for quasi-uniform k-space sampling.

For our RL-constrained cardiac MRI reconstruction framework, we introduce a hemisphere-adapted golden angle formulation:

\begin{equation}
\phi_g = \frac{\pi(2-\phi)}{2} = \frac{137.508°}{2} = 68.754°
\end{equation}

\textbf{Theoretical Justification:} The hemisphere constraint arises from our RL action space $\mathcal{A} = \{0, 1, \ldots, 179\}$, corresponding to angular domain $[0°, 180°]$. This adaptation exploits k-space Hermitian symmetry $S(-k_x, -k_y) = S^*(k_x, k_y)$, enabling complete reconstruction from hemisphere sampling while preserving the quasi-uniform distribution properties essential for optimal k-space coverage.
Within the constrained domain $[0°, 180°]$, the hemisphere-adapted increment $\phi_g$ maintains the discrepancy minimization properties of the classical golden ratio, ensuring uniform angular distribution with convergence rate $O(\log N/N)$ for $N$ sampling points.

The integration of this mathematically principled sampling strategy with our RL framework enables adaptive trajectory optimization while maintaining theoretical guarantees for k-space coverage uniformity.

\section{Reconstruction Network Specification}

Our framework employs a \textbf{direct inverse Fourier Transform (IFFT)} reconstruction approach using the FastMRI library's 2D implementation, followed by magnitude computation and min-max normalization. This design choice was strategically motivated by three key considerations: \textit{computational efficiency} for real-time reward computation during reinforcement learning training; \textit{methodological focus} for isolating sampling pattern optimization by eliminating reconstruction algorithm variables; and \textit{fair comparison} ensuring performance improvements are attributed to sampling strategies rather than reconstruction techniques.

While representing a performance lower bound compared to state-of-the-art reconstruction networks, this approach provides a principled baseline for evaluating learned sampling patterns without confounding factors. The orthogonal nature of our sampling optimization enables ready integration with any reconstruction algorithm to achieve synergistic improvements, ensuring our contributions remain transferable across the broader spectrum of MRI reconstruction techniques.

\section{Interpretation of Global Attention in Reward Function}

The "global attention" mechanism refers to the \textit{adaptive weighting coefficient} $\alpha$ that balances global image quality metrics against region-specific anatomical measures. The coefficient decreases linearly from 0.4 to 0.1 during training:

$$\alpha(t) = \max(0.1, 0.4 - 0.3 \cdot \frac{t}{T})$$

where $t$ and $T$ represent current and total training steps, respectively.

This decay strategy mimics clinical diagnostic workflows where radiologists first assess overall image quality before examining specific anatomical regions. The progressive shift from global ($\alpha$) to anatomical emphasis ($\beta = 1 - \alpha$) enables the agent to establish fundamental reconstruction competency before optimizing diagnostically relevant structures. A threshold-based mechanism ($|\text{SSIM}_{\text{global}} - \text{SSIM}_{\text{anatomical}}| < 0.01$) maintains balanced attention when metrics converge, preventing premature optimization.

\section{Multi-line to Single-line Acquisition Transition}

The transition from multi-line to single-line acquisition was driven by \textit{training dynamics optimization} based on empirical observations. Initially, we implemented two radial lines per action to provide substantial state changes and facilitate exploration during early training phases.

The switch to single-line acquisition was motivated by three factors: \textit{enhanced control granularity} for finer-grained sampling decisions and precise trade-offs between radial directions; \textit{improved training stability} as multi-line acquisition generated larger policy gradients causing occasional instability; and \textit{convergence optimization} through more stable gradient updates. 

\section{State-of-the-Art Reconstruction Context}

While our primary contribution focuses on sampling pattern optimization rather than reconstruction algorithm innovation, we acknowledge the importance of contextualizing our approach within contemporary radial MRI reconstruction methodologies. 
Our reinforcement learning-based sampling optimization represents a \textit{complementary approach} that is orthogonal to reconstruction algorithms. Using direct IFFT reconstruction establishes a performance lower bound, while the learned sampling patterns can enhance any reconstruction method through superior data acquisition strategies. Critically, our anatomically-aware reward function addresses a gap in existing methods that optimize global image quality without explicit consideration of diagnostically relevant regions. The sampling patterns learned through our approach can be directly integrated with state-of-the-art reconstruction networks, potentially achieving synergistic improvements through optimal data acquisition combined with advanced reconstruction algorithms.

\end{document}